\documentclass[11pt]{article}
\pdfoutput=1
\usepackage{jheppub}
\usepackage{hyperref}
\usepackage{graphicx,epstopdf,amsmath,amsfonts,amssymb,appendix,comment}
\usepackage{color,slashed,subfigure,setspace,footnote,multirow,longtable,braket}
\usepackage{epsfig,mathrsfs,latexsym,color,url,etoolbox}
\usepackage{bbm}
\definecolor{nicered}{rgb}{0.7,0.1,0.1}
\definecolor{nicegreen}{rgb}{0.1,0.5,0.1}
\definecolor{CarnationPink}{rgb}{1.0, 0.65, 0.79}
\DeclareMathAlphabet{\mathbbold}{U}{bbold}{m}{n}    

\usepackage{comment}

\usepackage{float}

\allowdisplaybreaks
\usepackage{dcolumn}
\usepackage{bm}
\usepackage[table]{xcolor}
\usepackage{multirow}
\usepackage{comment}
\usepackage{url}
\usepackage{tcolorbox}
\usepackage[T1]{fontenc}
\usepackage{cases}

\definecolor{nicegreen}{rgb}{0., 0.75, 0.46}

\usepackage{ulem}

\begin{document}

\preprint{PITT-PACC-2312, CERN-TH-2023-030, MI-HEE-801}

\title{{\huge Keeping it Simple:}\\ {\Large Simplified Frameworks for Long-Lived Particles at Neutrino Facilities}}

\author[1]{Brian Batell,}
\author[1]{Wenjie Huang,}
\author[2,3]{Kevin J. Kelly}
\affiliation[1]{Pittsburgh Particle Physics, Astrophysics, and Cosmology Center, Department of Physics and Astronomy, University of Pittsburgh, Pittsburgh, PA 15217, USA}
\affiliation[2]{Theoretical Physics Department, CERN, Esplande des Particules, 1211 Geneva 23, Switzerland}
\affiliation[3]{Department of Physics and Astronomy, Mitchell Institute for Fundamental Physics and Astronomy, Texas A\&M University, College Station, TX 77843, USA}
\emailAdd{batell@pitt.edu}
\emailAdd{weh68@pitt.edu}
\emailAdd{kjkelly@tamu.edu}

\date{\today}

\abstract{
Modern-day accelerator neutrino facilities are excellent venues for searches for new-physics particles. Many distinct new-physics models predict overlapping signatures and phenomenology in these experiments. In this work, we advocate for the adoption of \textit{simplified frameworks} when studying these types of new-physics signatures, which are characterized by a small number of primary variables, including particle masses, lifetimes, and production and decay modes/rates that most directly control signal event rates and kinematics. In particular, taking the example of long-lived particles that decay inside a neutrino detector as a test case, we study formulate and study simplified frameworks in the context of light scalars/fermions produced in kaon decays which then decay into final states containing an electron-positron pair. We show that using these simplified frameworks can allow for individual experimental analyses to be applicable to a wide variety of specific model scenarios. As a side benefit, we demonstrate that using this approach can allow for the T2K collaboration, by reinterpreting its search for Heavy Neutral Leptons, to be capable of setting world-leading limits on the Higgs-Portal Scalar model. Furthermore, we argue the simplified framework interpretation can serve as a bridge to model identification in the hopeful detection of a new-physics signal. As an illustration, we perform a first determination of the likelihood that, in the presence of a new-physics signal in a detector like the DUNE ND-GAr, multiple different new-physics hypotheses  (such as the Higgs-Portal Scalar and Heavy Neutral Lepton ones) can be disentangled. We demonstrate that this model discrimination is favorable for some portions of detectable new-physics parameter space but for others, it is more challenging.
}

\maketitle

\section{Introduction}\label{sec:Introduction}

The past decade has seen renewed exploration for weakly coupled new physics beyond the Standard Model (BSM) in the MeV-GeV mass range. Such light BSM particles frequently appear in models addressing some of the outstanding questions in particle physics (neutrino masses, dark matter, etc.), and in explanations of various experimental anomalies (the muon anomalous magnetic moment~\cite{Muong-2:2006rrc,Muong-2:2021ojo}, MiniBooNE low energy excess~\cite{MiniBooNE:2018esg,MiniBooNE:2020pnu}, etc.). Accelerator-based neutrino beam experiments offer a particularly powerful approach in the search for such new light degrees of freedom~\cite{Batell:2022xau}. In these experiments, the BSM particles can be copiously produced in the collisions of an intense proton beam with a fixed target and then leave a variety of striking signatures in a downstream near detector. In many cases, past neutrino beam experiments provide the leading constraints on light BSM particles (e.g., Refs.~\cite{CHARM:1985nku,Boiarska:2021yho}). Furthermore, modern neutrino beam experiments such as MiniBooNE~\cite{MiniBooNE:2017nqe,MiniBooNEDM:2018cxm}, ArgoNeuT~\cite{ArgoNeuT:2021clc}, MicroBooNE~\cite{MicroBooNE:2019izn,MicroBooNE:2021usw}, T2K~\cite{T2K:2019jwa}, COHERENT~\cite{COHERENT:2021pvd,COHERENT:2022pli}, and CCM~\cite{CCM:2021lhc,CCM:2021yzc,CCM:2021leg}
 are now carrying out dedicated searches for new light BSM states, and  experiments such as ICARUS~\cite{ICARUS:2004wqc}, SBND~\cite{MicroBooNE:2015bmn}, JSNS$^2$~\cite{JSNS2:2021hyk}, and DUNE~\cite{DUNE:2016hlj,DUNE:2021tad} will provide enhanced sensitivity to these scenarios in the near future and beyond.  
 Refs.~\cite{Batell:2022dpx,Gori:2022vri} (and references therein) provide a thorough summary of present and proposed searches for these states.

The design and interpretation of these new physics searches are often guided by the predictions that arise in certain theoretically well-motivated models. Popular examples are models based on the renormalizable neutrino, Higgs, and vector portals, as well as those employing the dimension 5 axion-like particle portals -- each of these is discussed in turn in Ref.~\cite{Batell:2022dpx}, demonstrating the current experimental constraints on each ``portal'' and future prospects for discovery. While this ``top-down'' approach is certainly warranted and should continue, here we would like to argue that there is also value in developing a more model-independent approach to BSM searches at neutrino experiments. In particular, in this paper we wish to introduce and advocate for the use of {\bf simplified frameworks} 
that are characterized by a handful of relevant quantities that most directly determine the event rates and final state kinematics of the signal under consideration. Depending on the signature under consideration, the quantities entering into the simplified framework description could include masses and lifetimes of particles, decay branching ratios, production and scattering cross sections, and production energy and position distributions. These primary quantities that are directly constrained or measured in a particular experimental analyses can be mapped to a variety of more complete descriptions in terms of simplified model Lagrangians, effective field theories, and ultraviolet complete models. This simplified framework also admits simple extension beyond well-studied models and allows for analyses to determine whether a prospective new-physics signal truly originates from one of these specific models or whether contributions from additional new physics are necessary. 
In fact, a similar approach has been adopted in an analysis of new-physics searches using Super-Kamiokande in Ref.~\cite{Coloma:2019htx} and MicroBooNE~\cite{MicroBooNE:2021usw,Coloma:2022hlv}, and, very recently for the ProtoDUNE experiment in Ref.~\cite{Coloma:2023adi}. Similarly, Ref.~\cite{Dobrich:2018jyi} proposed simplified frameworks for light pseudoscalars produced in B meson decays, and NA62 subsequently used these frameworks in their analysis~\cite{Dobrich:2023dkm}. See also Ref.~\cite{Costa:2022pxv} for a distinct approach to model agnostic new physics searches at neutrino experiments. 

There are several motivations for developing such an approach. 
From an experimental perspective, similar signatures involving the same detectable final state particles arise in a variety of distinct BSM models, motivating a more flexible theoretical framework. 
The presentation of experimental results in such a simplified framework, either in the case of null results/limits or in the advent of signal excess, would more readily allow for reinterpretations by theorists in a variety of complete models, including those models that have not yet been envisioned. 
Furthermore, a search that is optimized to cover a specific model may not provide the most effective coverage of other models predicting the same final state if the kinematics of the detected particles differ substantially. Instead, searches designed to maximize coverage with our proposed simplified framework may actually translate to a broader coverage of complete models due to the wider range of allowed final state kinematics. 
\begin{figure}
\begin{center}
\includegraphics[width=\linewidth]{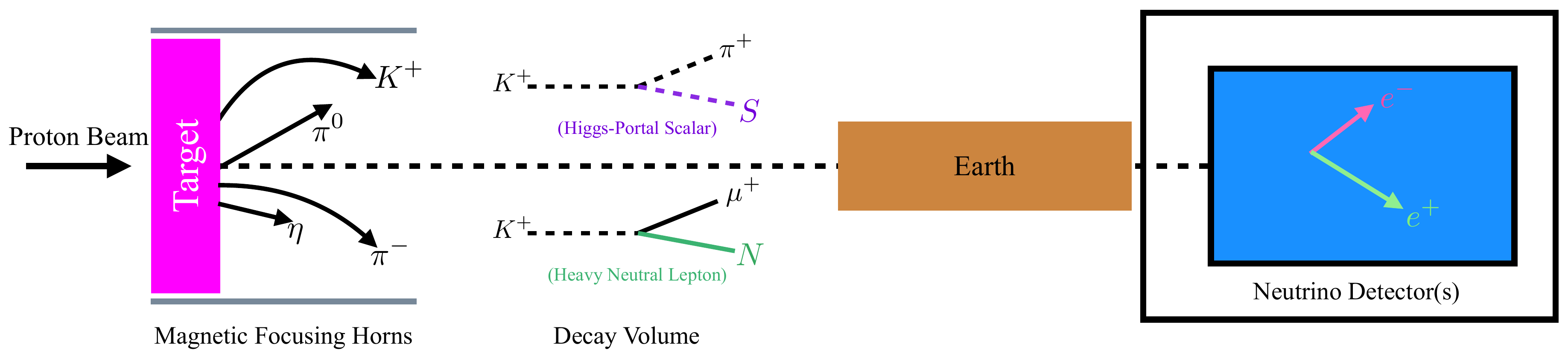}
\caption{Schematic outline of the processes considered in this work -- Standard Model mesons are produced in intense, proton-target collisions, some of which are focused in the beam direction. Rare decays of these mesons (we focus on charged kaons here) can produce new-physics particles, including well-studied model-specific ones such as Higgs-Portal Scalars (purple, $S$) and Heavy Neutral Leptons (green, $N$). These particles, if relatively long-lived, can decay inside of downstream neutrino near detectors into many different Standard-Model signatures. In this work, we focus on decays into electron/positron pairs. 
\label{fig:Schematic}}
\end{center}
\end{figure}

A wide variety of signatures of new light BSM states have been explored in recent years. In principle, a simplified framework approach can be developed for each of these signatures. In this work, in order to illustrate the general approach, we consider a simplified framework for long-lived particles (LLPs), focusing specifically on new BSM particles that are produced in meson decays and which decay in the detector to final states containing a detectable $e^+ e^-$ pair and possible invisible particles such as neutrinos (Fig.~\ref{fig:Schematic} provides an outline of what we consider here). The basic simplified framework is defined by a small number of parameters, including the mass of the BSM particle, its lifetime, and the branching ratios associated with the meson decay to the BSM particle and the BSM particle to the $e^+e^-$ final state. Within this framework, we consider two specific cases in which the new particle is a singlet scalar or fermion, which are both inspired by and, for certain choices of the branching ratios and lifetime, reproduce the well-motivated Higgs-Portal Scalar and Heavy Neutral Lepton models. We then discuss how existing experimental limits and the potential results of future searches at neutrino experiments may be interpreted within our simplified frameworks. We note that while our focus in this paper is on accelerator fixed-target neutrino beam experiments, it is likely that a simplified framework approach may be more generally useful for a variety of experiments/facilities searching for LLP signatures, including ATLAS and CMS~\cite{Alimena:2019zri}, LHCb~\cite{Craik:2022riw}, Belle-II~\cite{Belle-II:2018jsg}, NA62~\cite{NA62:2017rwk}, DarkQuest~\cite{Apyan:2022tsd}, FASER~\cite{FASER:2019aik}, FASER$\nu$~\cite{FASER:2019dxq}, SND@LHC~\cite{SNDLHC:2022ihg}, the Forward Physics Facility~\cite{Anchordoqui:2021ghd}, MATHUSLA~\cite{Curtin:2018mvb}, CODEXb~\cite{Aielli:2022awh}, and SHiP~\cite{Alekhin:2015byh}, among others.

We begin in Section~\ref{sec:SimplifiedModels} with a discussion of the simplified framework philosophy and a definition of two cases to be studied in this work. In Section~\ref{sec:Simulations} we review the basic features of several neutrino experiments and our approach to simulating the signatures predicted by our simplified framework. Section~\ref{sec:Sensitivity} surveys the existing experimental constraints and future experimental sensitivity to the framework. Section~\ref{sec:Measurement} considers the scenario in which an excess is observed at a future neutrino experiment and addresses how well one may measure simplified framework parameters and discriminate between the two cases studied. Finally, we present our conclusions and outlook for future simplified framework studies in Section~\ref{sec:Conclusions}.

\section{Common Features \& Simplified Frameworks for Long-Lived Particles}\label{sec:SimplifiedModels}
We begin by discussing the ideas that we can test at neutrino near-detector facilities, searching for decay signatures from LLPs. The steps for these signatures to occur manifest in many specific new-physics models that have been explored in this context to varying degrees. We wish to highlight the features shared across these models, which will then allow us to define the aspects of some simplified frameworks. These simplified frameworks, when explored properly, can both 
\begin{itemize}
\item be mapped onto model-dependent new-physics parameter spaces for sensitivity/exclusion and/or discovery, and
\item extend these model scenarios in new and non-trivial ways.
\end{itemize}

Let us consider the following scenario, a schematic representation of which is shown in Fig.~\ref{fig:Schematic}. Many modern-day and planned neutrino beams operate by 
impinging a proton beam
on a target to produce SM mesons which (after focusing for sign-selection) then decay into neutrinos. Many SM mesons (including neutral ones which rarely, if ever, decay into neutrinos) are produced, $\mathcal{O}(1)$ per proton-on-target (POT) for many different meson types.
New-physics scenarios predict, for specific mesons $\mathfrak{m}$, decays into a new particle $X$, of the type $\mathfrak{m} \to X + {\rm SM}$, where `SM' is one or more SM particles.
The new particle $X$ is often long-lived (on experimental scales) and can travel towards the nearby neutrino detector. 
Depending on the lifetime of $X$ and its various partial decay widths, it can decay into one or more SM final states $F$ via the process $X \to F$ in the detector. Often, the final states $F$ are notably distinct from SM backgrounds (e.g., from cosmic rays, neutrino scattering, etc.) in these detectors and these searches are sensitive to $\mathcal{O}$(few) events in their data collection period.

To be more exact, we can determine the relevant information that determines the signal rate in such a search. First, we determine the flux of particles $X$ entering the detector,
\begin{equation}\label{eq:PhiX}
\Phi_X = \frac{c_\mathfrak{m} N_{\rm POT}}{A_{\rm Det.}} \varepsilon\left(\mathfrak{m}; m_X,\dots\right) \mathrm{Br}\left(\mathfrak{m} \to X\right),
\end{equation}
where $c_\mathrm{m}$ is the $\mathcal{O}(1)$ number representing the number of mesons $\mathfrak{m}$ produced per POT at the given beam energy, $N_{\rm POT}$ is the number of POT delivered on the target, and $A_{\rm Det.}$ is the detector's transverse area as viewed by the incoming beam. $\varepsilon$ represents a geometrical efficiency factor, the fraction of particles $X$ 
produced in $\mathfrak{m}$ decays that emerge with three-momentum pointing towards the detector. This quantity is estimated using Monte Carlo simulation and can depend (beyond the $X$ mass) on the other SM particle(s) that emerge from the decay. Also, using simulation, we determine the energy ($E_X$) and spatial (along the beam axis, $z$) distribution of the $X$ particles that emerge from this decay, giving us the differential flux $d^2\Phi_X/dE_X dz$~\footnote{This will be sufficient for the experimental setups considered in this work featuring highly boosted kinematics and detectors that are small in extent in comparison to the baseline. The analysis can be straightforwardly generalized to account for the full momenta and spatial distributions as needed.}.
The signal rate of a final state $F$ is then
\begin{align}
N_{\rm sig.}^F = \int dE_X \int_{A_{\rm Det.}} dA \int_{0}^{z_{\rm max}} dz &\int_{0}^{L_{\rm Det.}} \left( \frac{d^2\Phi_X}{dE_x dz} P_{\rm Decay}\left(E_X, z' + D_{\rm Det.} - z\right) \mathrm{Br}\left(X \to F\right) \right) dz', \label{eq:NSig}\\
P_{\rm Decay}\left(E_X, \zeta\right) &= \frac{1}{\gamma_X \beta_{X} c\tau_X} e^{-\frac{\zeta}{\gamma_X \beta_{X} c\tau_X}};\quad \gamma_X = E_X/m_X
\end{align}
From this exercise, we can extract that, beyond experiment-specific information, the relevant factors that enter $N_{\rm sig.}^F$ are
\begin{itemize}
\item The SM meson $\mathfrak{m}$ of interest that decays into the new-physics particle $X$. Of importance is the branching ratio $\mathrm{Br}\left(\mathfrak{m} \to X\right)$.
\item The mass of the decaying particle $X$, $m_X$. This impacts the spectrum of $X$ particles emerging from $\mathfrak{m} \to X$ decays, as well as the boost $\gamma_X$.
\item The signal of interest $F$, including whether or not $F$ is a fully-visible final state or if some particle(s), such as neutrinos, emerge undetected. The quantity $\mathrm{Br}\left(X \to F\right)$ directly enters the signal rate.
\item The lifetime of $X$, $c\tau_X$.
\end{itemize}
Specifically, $N_{\rm sig.}^F$ depends only on the product $\mathrm{Br}\left(\mathfrak{m} \to X\right) \mathrm{Br}\left(X\to F\right)$. Once a specific parent meson $\mathfrak{m}$ and signal channel $F$ are chosen, then this simplified approach depends only on three parameters -- $m_X$, $c\tau_X$, and this branching-ratio-product. 
Whether or not the decay $X \to F$ contains invisible particles will modify the signal kinematic distributions of $F$ observed by the experiment but not the overall rate -- we will discuss this effect in detail in the coming sections.

We advocate for utilizing such simplified frameworks in neutrino near-detector experiments when searching for LLPs, as it can allow for these analyses to apply for a wider range of model-specific scenarios than analyses performed with a specific model in mind. In the event of a discovery, these simplified frameworks may serve as templates to determine whether the discovery is consistent with one well-studied minimal models (such as Heavy Neutral Leptons, Higgs-Portal Scalars, Dark Photons, etc.) or not. Using this approach can allow for determining if this new physics contains additional complications beyond these minimal model scenarios.

For specificity and to illustrate the general approach, we will choose one parent meson and one final-state signal channel for the remainder of this work, however, the conclusions we draw will be illustrative of many other choices as well. We opt to focus on $K^\pm$ decays into particles $X$ that can decay into electron/positron pairs. This choice is motivated in part by the observation that neutrino experiments sourced by ${\sim}10-100$ GeV proton beams can produce a large number of kaons, but not too many heavier mesons. Additionally, outside of signals with one or more photons, the signal $F \supset e^+e^-$ will allow us to search for the lightest possible $X$ particles. This combination of choices will allow us to consider $2m_e < m_X \lesssim m_{K^\pm}$, a fairly wide range accessible to these experiments. 

In particular, we will consider two distinct cases within the simplified framework with the above properties in what follows. The first case contains a long-lived neutral scalar $S$ which is produced via $K \rightarrow \pi S$ and decays via $S \rightarrow e^+ e^-$. This case is described by the parameters
\begin{align}
\label{eq:SimpleScalar}
m_S, ~~~~~ c \tau_S, ~~~~~~  {\rm Br}(K\rightarrow \pi S) \times {\rm Br}(S\rightarrow e^+ e^-)  ~~~~~~~~~~ {\rm (Scalar~Case) }
\end{align}
The second case we will study features a long-lived neutral fermion that is produced via $K\rightarrow \mu N$ and decays via $N\rightarrow e^+ e^- \nu$, which is described by the parameters
\begin{align}
\label{eq:SimpleFermion}
m_N, ~~~~~ c \tau_N, ~~~~~~  {\rm Br}(K\rightarrow \mu N) \times {\rm Br}(N\rightarrow e^+ e^- \nu)  ~~~~~~~~~~ {\rm (Fermion~Case) }
\end{align}
These are of course inspired by the Higgs-Portal Scalar and the muon-flavor mixed Heavy Neutral Lepton models, respectively, and can be directly mapped onto these models for certain choices of parameters. While both are produced through kaon decays and give the same final state, there are some important differences. Perhaps most importantly, the different decays of $S$ and $N$ lead to distinctive kinematics of the final state electron and positron. 
By studying the two cases above, we will be able to illustrate how a single experimental search (a $e^+ e^-$ final state in this work) can apply to a variety of models. It will also allow us to test how well experiments can discriminate between different models in the advent of a signal excess, a question which we will take up in Section~\ref{sec:Measurement}. Further interesting physics models, such as dark photons or axion-like particles, which decay into electron/positron final states, can be studied in similar frameworks that depend on different production mechanisms, and would depend on similar parameters to those in Eqs.~\eqref{eq:SimpleScalar}-\eqref{eq:SimpleFermion}.

\subsection{Example Mapping to Simplified Lagrangian, EFT, and UV Model}\label{subsec:Lagrangian}
Here we provide an example mapping of the simplified scalar case (\ref{eq:SimpleScalar}) to more theoretically complete descriptions.
At the first stage, we can introduce a simplified-model Lagrangian involving the scalar $S$ and SM particles involved in the production and decay of $S$:
\begin{align}
    \mathcal{L} &\supset -\frac{1}{2} m_S^2 S^2 - g_{K\pi} S \pi^- K^+ + \mathrm{h.c.} \nonumber \\
    &- g_e S \bar{e}e - g_{\chi} S \bar{\chi} \chi,\label{eq:ScalarL}
\end{align}
where $g_{K\pi}$ and $g_{e}$ are effective couplings which regulate kaon decay into $S$ and $S$ decay into electron/positron pairs, respectively. The final term, proportional to $g_\chi$, allows $S$ to decay into (undetectable) dark matter pairs (assuming such a decay is kinematically accessible). The relative size of $g_e$ and $g_\chi$ allows for the branching ratio into visible final states and the overall lifetime to be independent parameters.

In well-studied dark-sector-portal scenarios, the parameters of Eq.~\eqref{eq:ScalarL} can be mapped on to other model parameters. 
In the case of the minimal Higgs-Portal Scalar, which we denote by  
$\varphi$ to distinguish it from the simplified scalar $S$, the only interactions of $\varphi$ with the SM are via a small mixing with the SM Higgs parameterized by angle $\sin\vartheta$. We can determine this mapping (see, e.g., Ref.~\cite{Winkler:2018qyg}),
\begin{align}
 g_{K\pi}  & = \frac{3 m_t^2 V_{td}^* V_{ts}}{32 \pi^2 v^3} (m_K^2-m_\pi^2)\sin\vartheta \,, \label{eq:gKpiDef} \\
 g_e & = \frac{m_e}{v}\sin\vartheta \, ~  \label{eq:geDef}.
\end{align}
Here, $m_t$ ($m_e$) is the top quark (electron) mass, $v=246$ GeV is the Higgs vacuum expectation value, and $V_{td}$ and $V_{ts}$ are the relevant elements of the CKM matrix. Eq.~(\ref{eq:gKpiDef}) arises from determining the kaon decay width into a pion and scalar in the two scenarios, where in the simplified scenario (\ref{eq:ScalarL}) it is a tree-level two-body decay while for the Higgs-Portal Scalar, it is generated at loop level (hence the dependence on the top quark mass and CKM matrix elements).
Similarly, Eq.~(\ref{eq:geDef}) is obtained by relating the partial width of $S$ (or $\varphi$) into an electron/positron pair in the two scenarios. It is evident, comparing Eqs.~\eqref{eq:gKpiDef} and~\eqref{eq:geDef}, that within the Higgs-Portal Scalar model, the effective couplings $g_{K\pi}$ and $g_e$ are bound by this single parameter $\sin\vartheta$, forcing the widths $\Gamma(K \to \pi S)$ and $\Gamma(S \to e^+e^-)$ to be fundamentally linked. In contrast, the freedom of the simplified model allows $g_{e}$ and $g_{K\pi}$, as well as these widths, to be independent.

The simplified model Lagrangian (\ref{eq:ScalarL}) can be realized in a variety of UV completions besides the minimal Higgs portal model.  To illustrate this, first note that coupling $g_{K\pi}$ in Eq.~(\ref{eq:ScalarL}) would arise from the quark-level effective Lagrangian
\begin{align}
    \mathcal{L} &\supset  - [g_{ds} \, S \, \overline d_L s_R +{\rm h.c.}],
    \label{eq:Scalar-L-quark}
\end{align}
with 
\begin{equation}
g_{K\pi} = g_{ds} \langle \pi | \overline d_L s_R | K \rangle, ~~~~~~~~  |\langle \pi | \overline d_L s_R | K \rangle| = \frac{1}{2} \frac{m_K^2 - m_\pi^2}{m_s-m_d}.
 \end{equation}
Moving up one level to take account of the SM gauge symmetries, the $S$-quark coupling in Eq.~(\ref{eq:Scalar-L-quark}) and the $S$-electron coupling in Eq.~(\ref{eq:ScalarL}) are obtained from the following dimension five operators:
\begin{align}
    \label{eq:Scalar-EFT}
{\cal L} \supset -\frac{{(C_d)}_i^j}{\Lambda} S \, \overline Q_L^i \, H \, d_{R\,j} - \frac{{(C_e)}_i^j}{\Lambda} S \, \overline L_L^i \, H \, e_{R\,j} 
+{\rm h.c.}\, ,
\end{align}
where $i,j$ are generational indices.
In the fermion mass basis, the couplings ${(C_d)}_1^2$ and ${(C_e)}_1^1$ should be nonvanishing to realize the Lagrangian (\ref{eq:Scalar-L-quark}).
Finally, the gauge invariant effective Lagrangian (\ref{eq:Scalar-EFT}) can descend from a variety of UV completions.
As one example, consider a model with an additional scalar doublet $\Phi \sim ({\bf 1}, {\bf 2}, \tfrac{1}{2} )$ with Lagrangian 
\begin{align}
{\cal L} & \supset | D_\mu \Phi |^2 - M_\Phi^2 | \Phi |^2 + \cdots  \nonumber \\
& - [ (y_d')_i^j \, \overline Q_L^i \, \Phi \, d_{R\,j}  + (y_e')_i^j \, \overline L_L^i \, \Phi \, e_{R\,j}    - A S \, H^\dag \, \Phi   +{\rm h.c.} ],
\end{align}
where the ellipses denote other scalar potential interactions. Integrating out the scalar at tree level leads to the EFT operators~(\ref{eq:Scalar-EFT}) with 
\begin{equation}
{(C_d)}_i^j = \frac{(y_d')_i^j A}{M_\Phi^2}, ~~~~~ {(C_e)}_i^j = \frac{(y_e')_i^j A}{M_\Phi^2}.
\end{equation}

It is also straightforward to construct UV completions of the fermion simplified framework, e.g., based on theories with new scalar doublets or vector bosons.  
We also note that the patterns of couplings between the new scalar or fermion could be qualitatively different than in the minimal Higgs-Portal Scalar or Heavy Neutral Lepton models. For example, the new scalar could dominantly couple to specific fermion flavors, such as electrons, rather than in proportion to their mass, see e.g., Refs.~\cite{Batell:2017kty,Egana-Ugrinovic:2018znw}.

\section{Experimental Simulations}\label{sec:Simulations}

In this section, we describe the steps that we take to simulate LLP searches in several neutrino experiments. We focus on two experiments' existing searches, mapped into our simplified framework scenarios, the MicroBooNE experiment and its search for Higgs-Portal Scalars coming from kaon decay-at-rest (KDAR) in the Neutrinos from the Main Injector (NuMI) beam dump~\cite{MicroBooNE:2021usw}, as well as the Tokai-to-Kamioka (T2K) Near Detector 280 (ND280) and its search for Heavy Neutral Leptons~\cite{T2K:2019jwa}. We also project forward to the Deep Underground Neutrino Experiment (DUNE) and searches capable at its near detector complex, building off studies presented in Refs.~\cite{Berryman:2019dme,Kelly:2020dda,Dev:2021qjj}. The approach used by MicroBooNE is somewhat different than the setup shown in Fig.~\ref{fig:Schematic}, however the simplified frameworks still present a suitable analysis framework in this case. In fact, the MicroBooNE collaboration adopted such an approach in Ref.~\cite{MicroBooNE:2021usw}, presenting constraints as a function of lifetime and branching ratios for a variety of scalar masses.

Ordering these experiments in this way allows us to develop more simulation complexity in each step. In discussing Eqs.~\eqref{eq:PhiX} and~\eqref{eq:NSig}, we motivated the differential $X$ flux with respect to its energy and position of production. For the MicroBooNE KDAR set-up, all $X$ are produced at the same location with the same energy (assuming a two-body kaon decay). In our simulation of T2K, we extend this approach to include an energy distribution of the $X$ particles\footnote{As we discuss in the following, we do not include beamline $z$-dependent production of $X$ in our simulation of T2K. This yields conservative estimates for very short-lived $X$ compared to what a more complete simulation would determine.}, and in our DUNE simulation, we allow for $X$ production along throughout the decay volume (as sketched in Fig.~\ref{fig:Schematic} and described by Eq.~(\ref{eq:NSig})). We give details of each experimental simulation in the following subsections.

Between now and a full analysis with future DUNE data, studies of this nature can be carried out in the SBND detector~\cite{MicroBooNE:2015bmn}, using the booster neutrino beam (BNB) as a source. However, kaon production is smaller given the lower BNB beam energy relative to those from the NuMI, DUNE, and T2K beams, and additionally, neutrino-beam induced backgrounds that mimic electron/positron signatures may be significantly larger. The ICARUS detector~\cite{ICARUS:2004wqc}, using the NuMI KDAR strategy similar to MicroBooNE as well as kaons produced in the NuMI target and along the beamline, may be able to improve on current constraints given the large ICARUS detector volume.

\subsection{MicroBooNE KDAR}\label{subsec:MicroBooNEDetails}
Motivated by the proposal in Ref.~\cite{Batell:2019nwo}, the MicroBooNE collaboration performed a study searching for Higgs-Portal Scalars that are produced in $K^+$ decays inside the NuMI absorber. This yields an isotropic scalar flux which can traverse the ${\sim}100$ m from the absorber to MicroBooNE. The scalar can decay within, and MicroBooNE placed constraints on such decays into electron/positron pairs in Ref.~\cite{MicroBooNE:2021usw}. More recently, Ref.~\cite{Kelly:2021xbv} utilized these results to place a constraint on the scenario where a kaon decays into a heavy neutral lepton, which then decays into a neutrino \textit{and} this electron/positron pair in the detector. Both model scenarios exhibited improved constraints over existing ones for a variety of Higgs-Portal Scalar/Heavy Neutral Lepton masses.

We use the details of Ref.~\cite{MicroBooNE:2021usw} -- especially its mass-dependent reconstruction efficiency\footnote{This is provided as an ancillary file in the arXiv submission of Ref.~\cite{MicroBooNE:2021usw}.} -- to reproduce its results. The MicroBooNE analysis focuses on masses below $210$ MeV -- we extend this to masses just below the kaon mass by extrapolating the reconstruction efficiency, using ${\sim}9\%$ efficiency in the larger mass range. This analysis considered $1.93 \times 10^{20}$ POT and had very small backgrounds when training specifically for Higgs-Portal Scalars in the mass range of interest. It is unclear what the background rate would be in a more model-independent analysis and so we cannot determine how MicroBooNE sensitivity will continue to scale with increased exposure. ICARUS can likely improve on these constraints given its larger detector volume and even more beam exposure.

\subsection{T2K}\label{subsec:T2KDetails}

In Ref.~\cite{T2K:2019jwa}, the T2K Collaboration, using its near detector ND280, reports constraints on Heavy Neutral Leptons with various mixing angles $|U_{\alpha}|^2$ ($\alpha=e,\ \mu,\ \tau$) with masses $m_N$ between $140 - 493$ MeV. All of these are considered to be produced in kaon decays, either $K \to e N$ or $K \to \mu N$, and various final states are considered. In minimal Heavy Neutral Lepton models, the branching ratios into certain final states, for example $N \to \mu \pi$, are predicted to be significantly larger than others, including $N \to \nu e^+ e^-$, which is the focus of our study. Regardless, T2K provides signal reconstruction efficiencies for various final states, including for $N \to \nu e^+ e^-$ where the efficiency is approximately $10\%$ over the entire range of $m_N$ considered. For our T2K approximations, we will assume that this efficiency is $10\%$, even for $1$ MeV $\lesssim m_N < $ 140 MeV, outside the range analyzed in Ref.~\cite{T2K:2019jwa}, similar to the approach considered in Ref.~\cite{Arguelles:2021dqn}.

We also utilize the heavy neutrino flux distributions provided in Ref.~\cite{T2K:2019jwa}, giving us the spectrum of $d\Phi_X/dE_X$ to enter into Eq.~\eqref{eq:NSig}. We take the spectrum provided for massless $X$ (properly rescaled to mitigate model assumptions and to match the official T2K results) in our simulations -- this will yield a \textit{conservative} result, particularly when considering particles $X$ with mass close to the kaon mass. This is because, for heavier masses, a larger fraction of $X$ particles emerging from kaon decay will have sufficiently small transverse momentum that they are pointing towards the T2K ND280 near detector, increasing $\varepsilon(K^\pm;m_X)$. Effectively, we are taking this efficiency to be constant for all $m_X$ considered. Additionally, because we do not have a detailed simulation of the T2K focusing magnets, we also assume that all $X$ are produced in/near the T2K target. This will lead to a conservative estimate when constraining $X$ with lifetimes on the order of, or smaller than, the distance between the target and the near detector. Finally, our estimates for T2K are based on Ref.~\cite{T2K:2019jwa} which considered $12.34\times 10^{20}$ POT. Due to the very small background rates in T2K's gaseous time-projection chambers, we expect that the sensitivity to new $X$ particles will continue to scale inversely to the exposure in coming years.

\subsection{DUNE}\label{subsec:DUNEDetails}
To simulate DUNE sensitivity to these LLP searches, we make use of charged kaon distributions from the DUNE Beam Interface Working Group which account for the (anti)focusing of (negatively-)positively-charged mesons when operating in neutrino mode, and vice versa in antineutrino mode. This also allows us to keep track of where the kaons (hypothetically) decay into the new-physics particle $X$. By simulating the $K\to X$ decays using Monte Carlo, we are able to incorporate both the energy and position spectra of the $X$ flux, which allows us to include $d^2\Phi_X/dE_X dz$ in Eq.~\eqref{eq:NSig}. 

For our simulations, we focus on the proposed DUNE ND-GAr (gaseous argon time projection chamber) detector, situated at a distance $z \approx 579$ m with a length in the beam direction of approximately $5$ m. We do so due to the low-energy thresholds, low density, exquisite particle identification, and magnetic field (and therefore, charge identification) present in the ND-GAr detector, all of which combine to allow for practically zero-background searches for LLPs. Ref.~\cite{Berryman:2019dme} studied backgrounds for $e^+ e^-$ signals in DUNE ND-GAr in detail, demonstrating how the expected background of ${\sim}5\times 10^{5}$ NC$\pi^0$ events (where $\pi^0 \to \gamma\gamma$ decay occurs in the GAr TPC) may be reduced to ${\sim}10^3$ background events by cutting on features such as the hadronic activity and the presence of two photons interacting in the ECAL, which do not reduce the signal rate at all. Subsequent cuts may reduce that background to $\mathcal{O}(0.1 - 1)$ leveraging the angular resolution of the detector (backgrounds will be more-or-less isotropic whereas the signal $e^+ e^-$ pair will be exactly or nearly in the beam direction). A detailed study is required to assess the optimal angular cut (and other potential cuts) given the specific signal in mind.

Additionally, such an analysis may also be performed with ND-LAr, the liquid-argon time projection chamber situated at $z \approx 574$ m, although background reduction (especially when considering electron/positron pairs) may prove more difficult. The sensitivity we obtain for DUNE will not depend significantly on this choice, as long as reasonable signal efficiencies and background reductions can be attained. 

Concretely, we assume five years of data collection using ND-GAr and the DUNE neutrino beam, with $1.47 \times 10^{21}$ POT per year. For sensitivity, we conservatively assume that ten signal events at DUNE will correspond to a stastically significant discovery.

We find that accounting for the $z$-distribution of the LLPs in our DUNE simulation improves sensitivity by an $\mathcal{O}(1)$ amount for proper lifetimes $c\tau_X \lesssim 1$ m -- if we assumed that all short-lived $X$ are produced at $z = 0$, then most/all would decay before reaching the DUNE near detector complex at $z \approx 574$ m. Allowing for production all the way until the end of the decay volume ($z \approx 230$ m) extends the sensitivity to significantly lower lifetimes.

\section{Simplified Framework Constraints \& Sensitivity}\label{sec:Sensitivity}
In this section, we demonstrate how the constraints from T2K and MicroBooNE discussed in Section~\ref{sec:Simulations} apply to the simplified frameworks we have introduced, as well as how DUNE will be sensitive to the same models. As laid out in Section~\ref{sec:SimplifiedModels}, we focus on three parameters -- the product $\mathrm{Br}\left(K\to X\right)\times \mathrm{Br}\left(X \to e^+ e^-\right)$, the mass of the LLP $m_X$, and its lifetime $c\tau_X$.

\begin{figure}
    \centering
    \includegraphics[width=\linewidth]{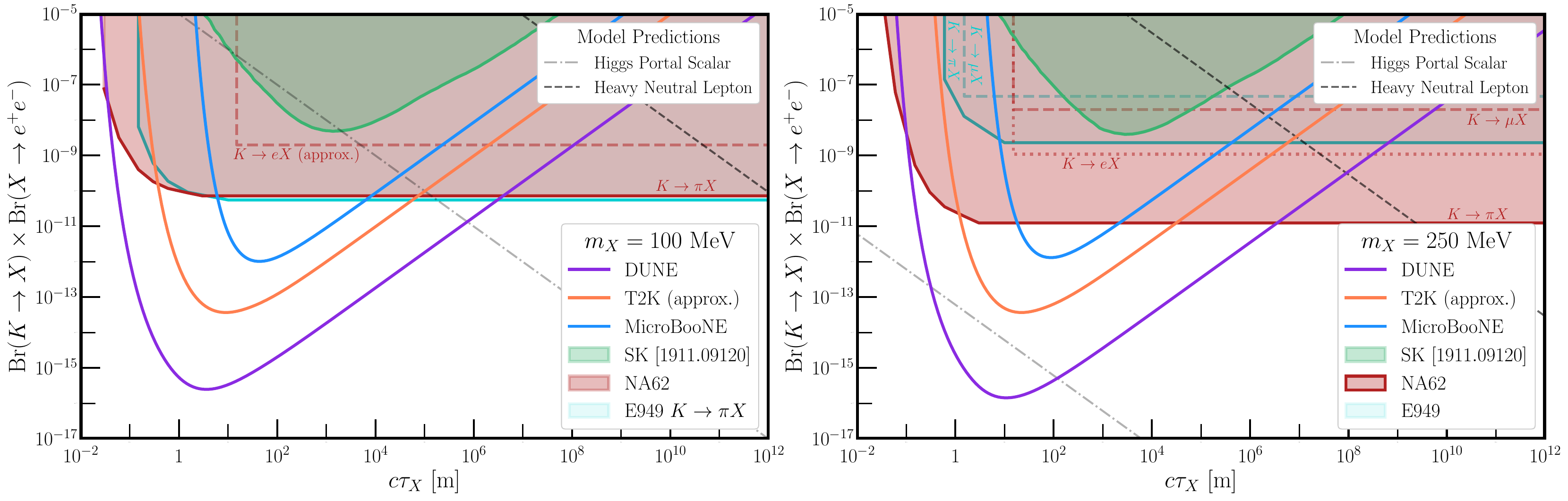}
    \caption{Simplified framework constraints and sensitivity as a function of LLP $X$ lifetime and the product of branching ratios Br($K\to X$)Br($X\to e^+e^-$) for two chosen $X$ masses, 100 MeV (left) and 250 MeV (right). We compare existing constraints from Super-Kamiokande~\cite{Coloma:2019htx} (green), E949~\cite{BNL-E949:2009dza,E949:2014gsn} (cyan), and NA62~\cite{NA62:2020mcv,NA62:2021bji,NA62:2020xlg,NA62:2021zjw} (red) with our recasts of existing constraints from T2K~\cite{T2K:2019jwa} (blue) and MicroBooNE~\cite{MicroBooNE:2021usw} (orange). Future projections utilizing the DUNE Near Detector are shown in purple. The dashed black and dot-dashed grey lines indicate predictions according to the Heavy Neutral Lepton and Higgs-Portal Scalar models, respectively.\label{fig:SimplifiedLifetimeBr}}
\end{figure}

First, we study how well the branching-ratio-product can be constrained as a function of $c\tau_X$ for several fixed values of $m_X$. The results of this process are displayed in Fig.~\ref{fig:SimplifiedLifetimeBr}. The neutrino-near-detector constraints (MicroBooNE in blue and T2K in orange) or sensitivity (DUNE in purple) all take a similar ``swoosh'' shape in this parameter space, where the left (right) panel displays constraints/sensitivities for $m_X = 100$ MeV ($250$ MeV).
In both panels, we compare against existing constraints from Super-Kamiokande in green (from Ref.~\cite{Coloma:2019htx}, which also explored this lifetime vs. branching-ratio-product in a simplified-framework context) and from NA62 in red~\cite{NA62:2020mcv,NA62:2021bji,NA62:2020xlg,NA62:2021zjw} and E949 in cyan~\cite{BNL-E949:2009dza,E949:2014gsn}. For both choices of $m_X$, we find that T2K presents stronger constraints for all $c\tau_X$ than MicroBooNE, and DUNE will surpass both. Model predictions for the Heavy Neutral Lepton assumption (dashed black) or Higgs-Portal Scalar (dot-dashed grey) are shown in each panel as lines. Some care is necessary in interpreting these constraints in model-specific scenarios -- we return to this discussion in Section~\ref{subsec:ModelInterp}.

In comparing the two panels of Fig.~\ref{fig:SimplifiedLifetimeBr}, we find compelling reasons for using these simplified frameworks when considering neutrino facilities for LLP searches. Notably, in comparing the curves from MicroBooNE, T2K, and DUNE (as well as the Super-Kamiokande constraint which, despite exploring atmospheric LLP production instead of beam-based, shares many of the features explored here) between the left and right panels, the parameter space constrained is very similar. This is expected as long as we focus on $m_X$ away from any kinematical endpoints (e.g. close to the kaon mass) where sensitivities will vary significantly. However, the model predictions move in this parameter space significantly, explaining why, in the model-specific parameter spaces, these seemingly-similar searches can appear very different.

\begin{figure}
    \centering
    \includegraphics[width=\linewidth]{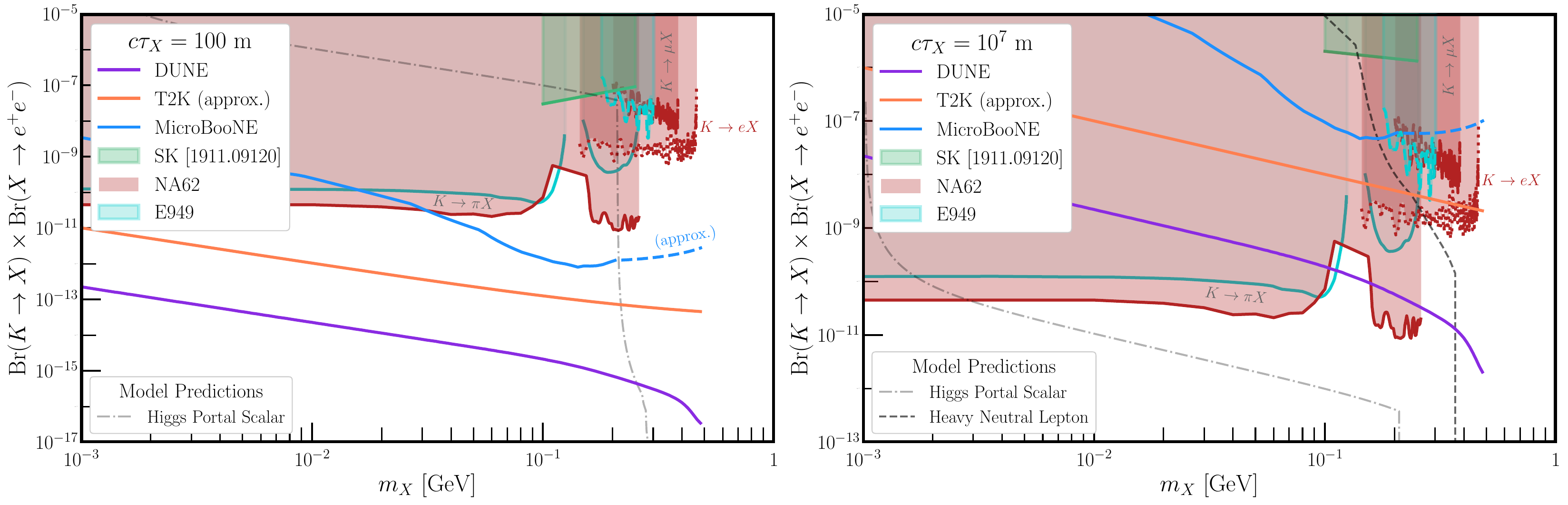}
    \caption{Simplified framework constraints and sensitivity as a function of LLP $X$ mass and the product of branching ratios Br($K\to X$)Br($X\to e^+e^-$) for two chosen $X$ lifetimes, 100 m (left) and $10^7$ m (right). We compare existing constraints from Super-Kamiokande~\cite{Coloma:2019htx} (green), E949~\cite{BNL-E949:2009dza,E949:2014gsn} (cyan), and NA62~\cite{NA62:2020mcv,NA62:2021bji,NA62:2020xlg,NA62:2021zjw} (red) with our recasts of existing constraints from T2K~\cite{T2K:2019jwa} (blue) and MicroBooNE~\cite{MicroBooNE:2021usw} (orange). Future projections utilizing the DUNE Near Detector are shown in purple. The dashed black and dot-dashed grey lines indicate predictions according to the Heavy Neutral Lepton and Higgs-Portal Scalar models, respectively.\label{fig:SimplifiedMassBr}}
\end{figure}

Fig.~\ref{fig:SimplifiedMassBr} similarly shows constraints and future sensitivity as a function of the branching-ratio-product, now as a function of $m_X$ for a fixed value of $c\tau_X$ -- the left (right) panel assumes $c\tau_X = 100$ m ($10^7$ m). We show the same set of constraints here as in Fig.~\ref{fig:SimplifiedLifetimeBr} -- Ref.~\cite{Yamazaki:1984vg} provides another constraint in this parameter space that is relatively weak compared to the other existing constraints from precision kaon deacy measurements. We see here that for different points in (mass, lifetime, branching-ratio-product) parameter space, different experiments excel relative to one another. In the left panel, the only model prediction shown is for the Higgs-Portal Scalar -- for a Heavy Neutral Lepton to have such a short lifetime, the branching ratio of kaons into the new particle must be extremely large, above the range shown in this figure.
Finally, we note again here (see Section~\ref{sec:Simulations} for more detail) that for MicroBooNE, above $m_X = 210$ MeV, we extrapolate the quoted efficiencies (hence the dashing of the blue line in Fig.~\ref{fig:SimplifiedMassBr}). For T2K, we take a constant 10\% efficiency for all masses.

\begin{figure}
    \centering
    \includegraphics[width=\linewidth]{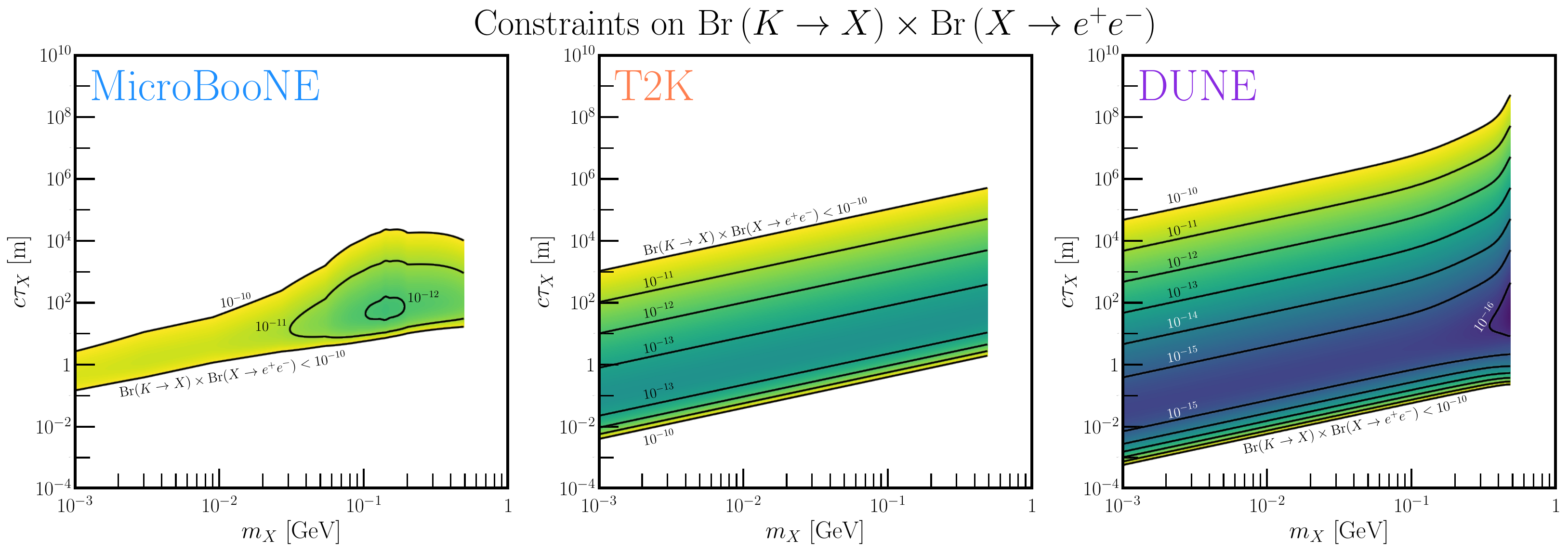}
    \caption{Contours of constraints on (or sensitivity to) the product $\mathrm{Br}(K\to X)\mathrm{Br}(X \to e^+e^-)$ for MicroBooNE (left), T2K (center), and DUNE (right) as a function of LLP mass and lifetime. Colored regions indicate where these constraints are stronger than $10^{-10}$, the (rough) constraint set across parameter space by NA62 and E949. Each successive black line indicates an order of magnitude smaller branching-ratio-product, as labelled.\label{fig:SimplifiedMass3D}}
\end{figure}

The final comparison we make for experimental sensitivity in these simplified-framework parameter spaces is shown in Fig.~\ref{fig:SimplifiedMass3D}, where we show as a contour plot, the sensitivity to the product $\mathrm{Br}(K\to X)\mathrm{Br}(X \to e^+e^-)$ for each of our simulations (left: MicroBooNE, center: T2K, right: DUNE) as a function of the LLP mass and lifetime. In all three panels, we shade the region for which each experiment constrains (or in the case of DUNE, is sensitive to) products of branching ratios below $10^{-10}$, roughly the strongest constraints set by NA62 and E949.

The shapes of these contours can be understood from our setup of Eq.~\eqref{eq:NSig}, emphasizing the power of using these simplified frameworks for LLP searches in neutrino facilities. For a given value of $m_X$, there is some $c\tau_X$ for which sensitivity is strongest, such that the probability of $X$ decaying inside the detector volume is largest. This occurs roughly when $\gamma c\tau_X \approx L_{\rm Det.}$, the distance between the bulk of the $X$ production and the detector. Note the factor of $\gamma$, the boost, here -- for heavier $m_X$, the distribution will have a characteristically smaller boost, implying that the sensitivity is optimized for larger $c\tau_X$. This explains the upward slope most notable in the center panel of Fig.~\ref{fig:SimplifiedMass3D} for T2K. The MicroBooNE constraints in the left panel deviate from this simple behavior because of the mass-dependent efficiency that we have included, which is highest at $m_X \approx 140$ MeV (providing the small island where MicroBooNE reaches $10^{-12}$ sensitivity). The final visible effect occurs close to threshold, where the simplified frameworks have a weakness. In our DUNE simulation (where we have explicitly simulated the $X$ production assuming $K^\pm \to e^\pm X$), as $m_X \to m_K$, the decay kinematics are such that a larger fraction of $X$ are pointing towards the near detector, enhancing $\Phi_X$. This leads to the ``expansion'' of the DUNE sensitivity region in the right panel of Fig.~\ref{fig:SimplifiedMass3D}.

\subsection{Model-Specific Interpretation Using Simplified-Framework Parameter Space}\label{subsec:ModelInterp}
\begin{figure}
    \centering
     \includegraphics[width=0.6\linewidth]{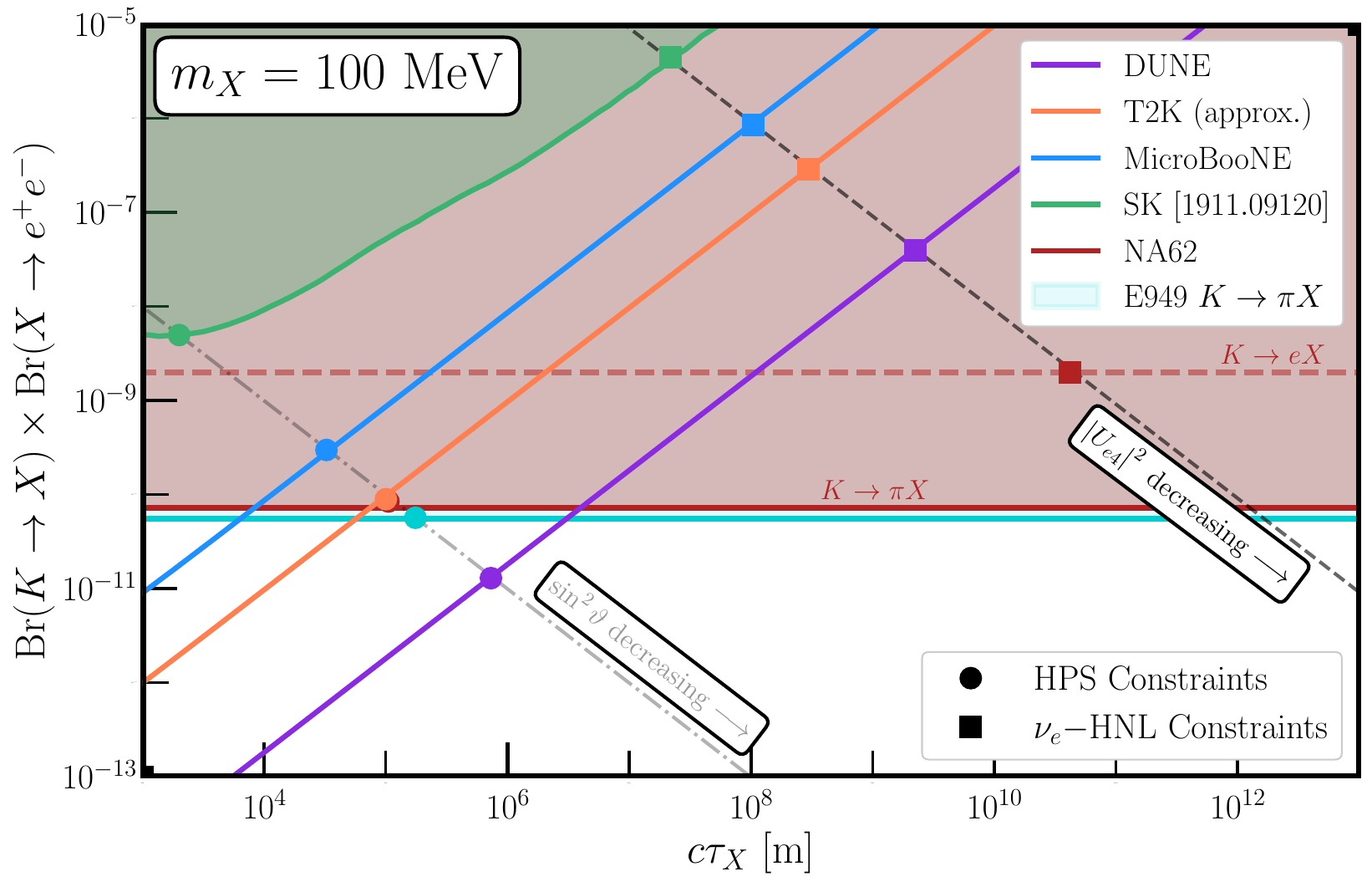}
    \caption{Schematic of how to place specific model constraints using simplified frameworks. When interpreting model constraints for LLP searches at neutrino experiments, the intersection of sensitivity/constraint curves with the model predictions determines the constraint on the relevant coupling ($\sin^2\vartheta$ for Higgs-Portal Scalar, $|U_{e4}|^2$ for Heavy Neutral Lepton). However, care (see text for details) is necessary for interpreting precision kaon-experiment constraints (e.g. NA62 and E949). \label{fig:ModelConstraints}}
\end{figure}
As discussed surrounding Figs.~\ref{fig:SimplifiedLifetimeBr} and~\ref{fig:SimplifiedMassBr}, we may project model-specific predictions (e.g. for the Heavy Neutral Lepton and Higgs-Portal Scalar models) onto the simplified frameworks when comparing experimental constraints and sensitivity because the lifetimes/predicted branching-ratio products are correlated for a given Heavy Neutral Lepton/Higgs-Portal Scalar mass. However, particularly when comparing neutrino-experiment sensitivities against those from other techniques (i.e., direct kaon-decay searches), some care is required. In this subsection, we explain how the simplified frameworks may be used for this model-depending constraint extraction.

Fig.~\ref{fig:ModelConstraints} presents a subset of the parameter space shown in Fig.~\ref{fig:SimplifiedLifetimeBr}(left), focusing on where different constraints/sensitivities intersect with the predictions from Higgs-Portal Scalar and Heavy Neutral Lepton models. As labelled in Fig.~\ref{fig:ModelConstraints}, changing the effective coupling in each model (the Higgs-Portal Scalar mixing $\sin^2\vartheta$ or the Heavy Neutral Lepton mixing angle $|U_{e4}|^2$) moves in this parameter space, where smaller couplings implies larger lifetimes and smaller branching ratios. If one wishes to interpret the neutrino-facility constraints (including Super-Kamiokande, MicroBooNE, T2K, and DUNE) in this parameter space with respect to a given model, then the intersection of the two lines determines the constraint set, as shown by the colored, filled circles and squares. However, the kaon-decay constraints (NA62 and E949) are distinct, as these searches rely on tagging the outgoing SM particle $Y$ in the final state $K \to X Y$. Moreover, analyses and constraints can vary significantly depending on $Y$, mostly due to SM backgrounds with certain final states. In Fig.~\ref{fig:ModelConstraints}, the constraints from NA62 differ by approximately one order of magnitude depending on whether outgoing pions (solid red line) or electrons (dashed red line) are considered. In the Higgs-Portal Scalar model, the outgoing particle in the kaon decay, along with $X$, is a pion, so the solid red and cyan lines (from NA62 and E949) intersecting with the grey dot-dashed line place a constraint on $\sin^2\vartheta$. However, for the Heavy Neutral Lepton model, the outgoing particle is a charged lepton (in this case, we focus on mixing via $|U_{e4}|^2$ so an outgoing electron). So, when placing constraints on $|U_{e4}|^2$, the intersection\footnote{Another caveat applies here -- NA62 (and other precision kaon-decay experiments) are sensitive just to $\mathrm{Br}(K\to eX)$, not the product with $\mathrm{Br}(X \to e^+e^-)$. In the Heavy Neutral Lepton scenario at $100$ MeV, the latter is predicted to be $\mathcal{O}(10^{-2})$, so the extracted constraint on $|U_{e4}|^2$ must be adjusted somewhat.} of the red dashed line with the black dashed line is what is important. We note here that the dashed red line is an extrapolation of the NA62 search for rare $K\to eX$ decays; neither NA62 or E949 have placed official constraints for masses near $m_X \approx 100$ MeV, hence there are no corresponding constraints on $|U_{e4}|^2$ in this mass range from these experiments.

\subsection{Updated Constraints on Higgs-Portal Scalar Parameter Space}
As discussed in Section~\ref{subsec:MicroBooNEDetails}, the MicroBooNE collaboration~\cite{MicroBooNE:2021usw} performed a dedicated search for the process $K^\pm \to \pi^\pm \varphi$, $\varphi \to e^+e^-$, where the kaon decay occurs in the NuMI absorber, the $\varphi$ travels from there to the MicroBooNE detector, and decays within. The same process can be studied by the T2K collaboration, with kaon production in the proton beam target, decay into $\varphi$ in the decay volume, and $\varphi$ decay into electron/positron pairs in ND280. While this specific process has not been studied by T2K, its search for Heavy Neutral Leptons~\cite{T2K:2019jwa} allows for extrapolation to this scenario, using the techniques detailed in Section~\ref{subsec:T2KDetails}.

If we compare the constraints in simplified framework parameter space across Figs.~\ref{fig:SimplifiedLifetimeBr} -~\ref{fig:SimplifiedMass3D}, we find that, for all combinations of LLP masses and lifetimes accessible by kaon production and with electron/positron final states, T2K-ND280 is able to constrain smaller values of $\mathrm{Br}(K \to X)\times\mathrm{Br}(X\to e^+e^-)$ than MicroBooNE-KDAR. Inspired by this, we turn from the simplified-framework parameter space to that of the Higgs-Portal Scalar, with mass $m_\varphi$ and mixing with the SM Higgs of $\sin\vartheta$.

MicroBooNE demonstrated world-leading constraints on this parameter space for $120$ MeV $\lesssim m_\varphi \lesssim 160$ MeV~\cite{MicroBooNE:2021usw}.
\begin{figure}
    \centering
    \includegraphics[width=0.5\linewidth]{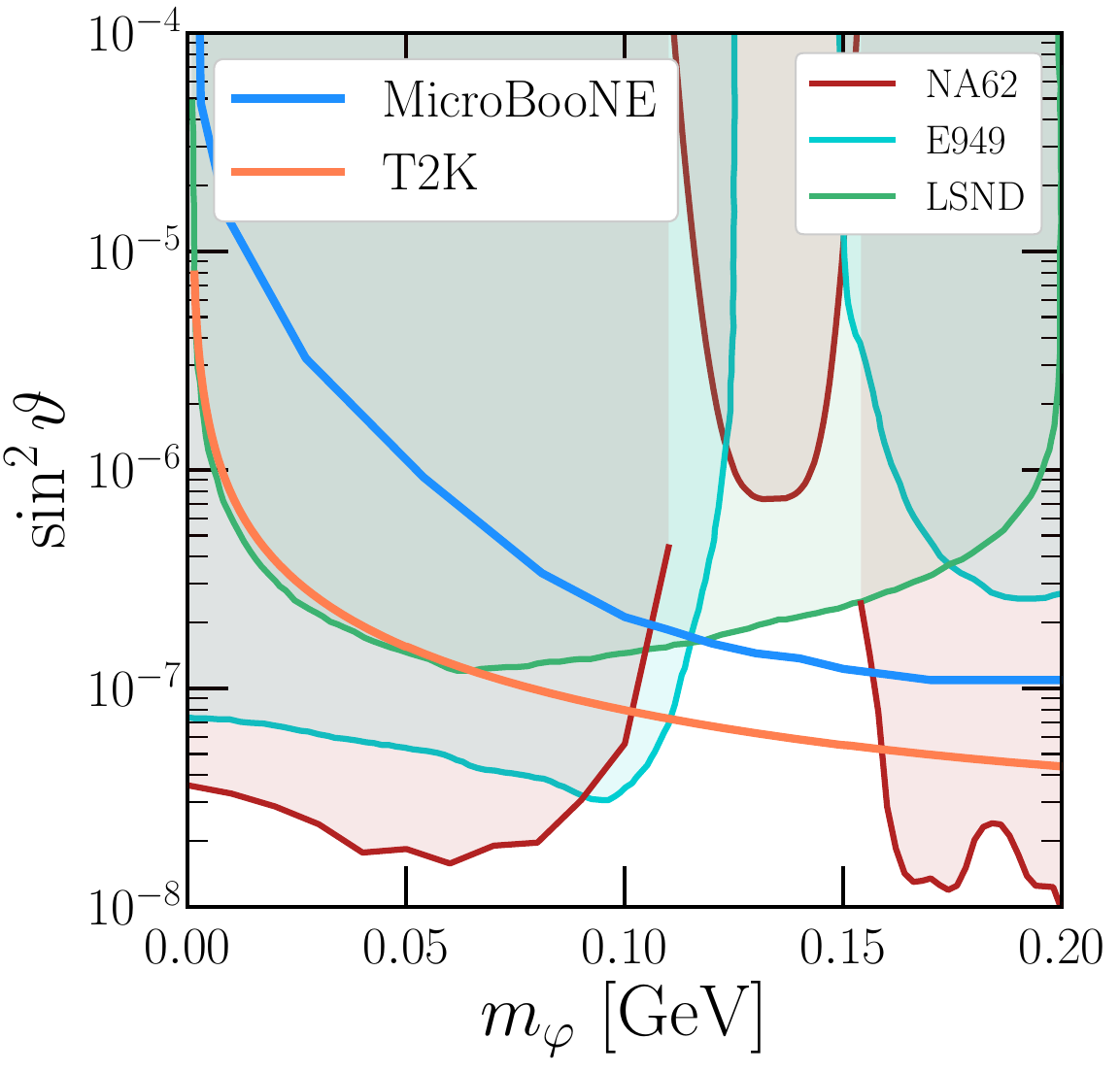}
    \caption{Constraints on a Higgs-Portal Scalar, using the simplified framework techniques of this work and mapped onto this specific parameter space for MicroBooNE (blue) and T2K (orange). For comparison, we show the strongest constraints on $\sin^2\vartheta$ as a function of $m_\varphi$, coming from NA62~\cite{NA62:2020pwi,NA62:2021zjw} (red), E949~\cite{BNL-E949:2009dza} (cyan), and LSND~\cite{Foroughi-Abari:2020gju} (green). See also Ref.~\cite{Gorbunov:2021ccu} for competitive bounds from the experiment PS191.\label{fig:DH_T2K_MicroBooNE}}
\end{figure}
When we apply our T2K simulation to this parameter space, we find that T2K improves on this MicroBooNE constraint by a factor of a few in terms of $\sin^2\vartheta$ across a wide range of masses, as presented in Fig.~\ref{fig:DH_T2K_MicroBooNE}. Here, we compare against existing constraints from NA62~\cite{NA62:2020pwi,NA62:2021zjw}, E949~\cite{BNL-E949:2009dza}, and LSND~\cite{Foroughi-Abari:2020gju}, which comprise the strongest constraints on $\sin^2\vartheta$ for this range of $m_\varphi$.

\section{Measurement in the Presence of Signal}\label{sec:Measurement}
In this section, we consider a possible future scenario in which a signal-event excess in a LLP search is observed at DUNE ND-GAr and discuss the interpretation of such an excess within our simplified framework. 
In particular, we attempt to answer the following two questions:
\begin{enumerate}
    \item how well the properties of a LLP can be measured, and
    \item how well the discrimination can be made between fully-visible final states and partially visible final states.
\end{enumerate}
As in the previous sections, 
we focus on the signature of a $e^{+}e^{-}$ final state of the scalar and fermion cases through a simplified framework approach. 

We assume a total number of 100 signal events in the $e^{+}e^{-}$ final states at DUNE ND-GAr in our simulation -- Fig.~\ref{fig:SimplifiedLifetimeBr} demonstrated the abundance of unexplored parameter space in which DUNE could observe 100 or more signal events.
As described in Section~\ref{sec:SimplifiedModels}, the simplified frameworks we advocate 
are characterized by three parameters - the branching ratio product $\mathrm{Br}\left(K\to X\right) \mathrm{Br}\left(X \to e^+ e^-\right)$, the mass of the LLP $m_X$, and its lifetime $c\tau_X$. 
The $e^+ e^-$ final state kinematics are sensitive to the LLP mass and lifetime, but not the branching ratio product. Therefore, we will provide interpretations of the excess by presenting our results as fits in the $m_X$-$c\tau_X$ plane, such that the fixed total number of 100 signal events is obtained by an appropriate choice of the branching-ratio product for each $(m_X,c\tau_X)$ point.

The following subsections detail the approach we take in simulating such a 100-event signal (Section~\ref{subsec:SimSetup}), and then in Sections~\ref{subsec:SimResults1} and~\ref{subsec:SimResults2}, we present results that address the two questions above regarding measurement capability and model discrimination.

\subsection{Simulated Event Distributions}\label{subsec:SimSetup}
Regardless of which question we are attempting to address, we must consider the observable event signatures in a detector like DUNE ND-GAr and how measurements of these kinematic observables can lead to various model discrimination power. 
We simulate data, containing information about the final-state particles, by first generating a large number of $K \to X$, $X \to e^+ e^- (\nu)$ decays. When reconstructing the electrons in the final state, we assume a $3^\circ$ angular resolution and a $5\%$ energy resolution on the final-state four-momenta, as a reference value~\cite{DUNE:2021tad}.
In reality, the energy and angular resolution will depend on the true energies and opening angles of an electron/positron pair, e.g. electron/positron pairs with smaller opening angles may be more challenging to measure due to possible overlap in the detector.

We then determine kinematical variables of interest from these reconstructed four-momenta, focusing on the following three:
\begin{itemize}
    \item The total energy $E_{e^+ e^-} = E_{e^+} + E_{e^-}$ of the electron/positron pair,
    \item The invariant mass $m_{e^+ e^-}$ of the electron/positron pair,
    \item The opening angle between the electron and positron emerging from the decay, $\theta_{e^+ e^-}$.
\end{itemize}
The simulated data are then binned into three-dimensional histograms with respect to these variables, with 20 bins in each dimension.\footnote{The range of the bins is chosen to include over 99\% of simulated events for comparison with other choices of parameters. We have found that the results of our analyses do not depend strongly on the number of bins chosen as long as it is larger than ${\sim}$10.}
As stated above, we assume that there are 100 signal events and so the histogram is then normalized. These variables are chosen to extract the maximum measurement potential and 
model discrimination power in the following ways:
\begin{itemize}
    \item The invariant mass $m_{e^+ e^-}$ is a good proxy for the parent mass $m_X$. In the fully visible decay case, $m_{e^+ e^-} \neq m_X$ due to uncertainties in measuring the outgoing electron/positron energy and direction. In the partially invisible case $N \to e^+ e^- \nu$, then $m_{e^+ e^-}$ will naturally be smaller than $m_N$ (within detector resolution effects). 
    \item The energy distribution $E_{e^+ e^-}$ serves as a way of measuring the lifetime $c\tau_X$. Depending on whether the $X$ is short- or long-lived, the distribution will tend to prefer larger or smaller $E_{e^+ e^-}$ so that the $X$ particles are able to reach the detector and decay within.
    \item The opening angle $\theta_{e^+ e^-}$ tends to be different depending on whether the decay is fully visible or partially invisible. When comparing these two hypotheses, including this kinematical variable assists in model discrimination. We also note that the angle of the $e^+ e^-$ pair relative to the beamline direction may also be able to shed light in terms of model discrimination, although we have not explored this potential in this work.
\end{itemize}

\begin{figure}
    \centering
    \includegraphics[width=\linewidth]{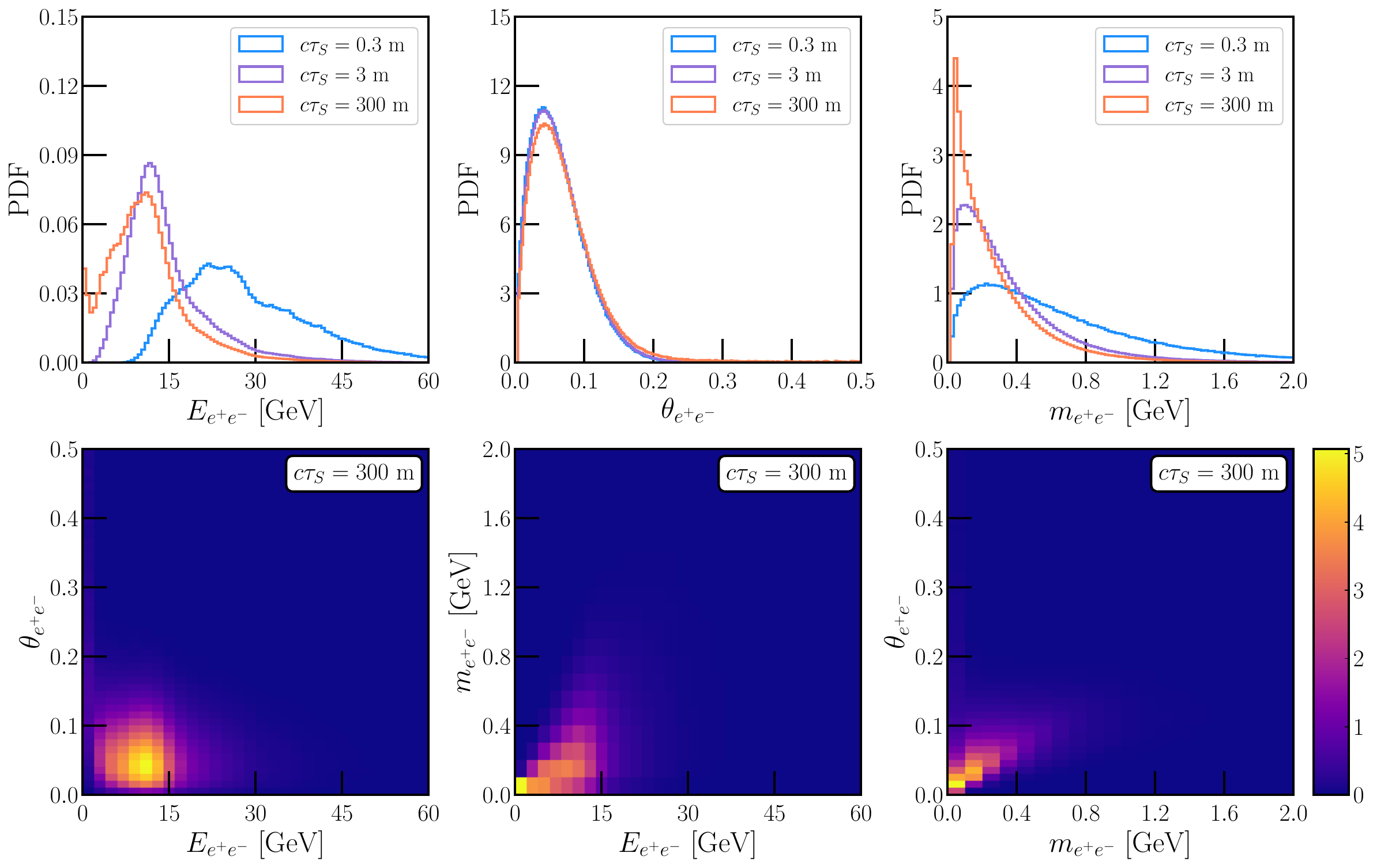}
    \caption{Event kinematics for scalars with mass $m_S = 50$ MeV decaying ($S \to e^+ e^-$) inside the DUNE ND-GAr detector. The top panels show one-dimensional kinematic distributions for the total electron/positron energy $E_{e^+ e^-}$ (left), the opening angle between the electron/positron pair (center), and the electron/positron pair invariant mass (right), for three different scalar lifetimes: 0.3 m (blue), 3 m (purple), and 300 m (orange). Two-dimensional distributions for each pair of kinematic variables are shown along the bottom row for the 300 m lifetime case.}
    \label{fig:KinDists}
\end{figure}

The distributions of these three kinematic variables are shown in Fig.~\ref{fig:KinDists} for a fully-visible decaying scalar $S \to e^+e^-$ with a mass of 50 MeV.
The bottom row gives the distribution in terms of pairs of kinematic variables for a scalar lifetime of 300 m. The top row gives the individual one-dimensional kinematic distributions for three choices of the scalar lifetime -- a very short lifetime of 30 cm (blue), a moderate lifetime of 3 m (orange), and a very long one of 300 m (purple). 
Notably, the short-lifetime of 30 cm in the top-left panel prefers significantly larger $E_{e^+ e^-}$ so that the scalars with high energy survive until reaching the detector instead of decaying beforehand. Because we have a 5\% relative energy measurement in our simulations, this means that those electrons/positrons are harder to measure accurately, leading to a broader distribution in $m_{e^+ e^-}$ (top-right panel), giving a distribution that no longer peaks at $m_S$. We see here that there are various features among the three distributions (and their correlations, only presented for $c\tau_S = 300$ m) that allow for discrimination between the three points in $\{m_S, c\tau_S\}$ parameter space.

In order to quantify the measurement potential or model discrimination for an assumed-true model, we compare the three-dimensional histograms of the truth and a test point using a $\chi^2$ test statistic (using Poissonian statistics for the possibly low bin counts),
\begin{equation}\label{eq:LogLikelihood}
\chi^2 \equiv -2\ln{L(m_X, c\tau_X)} = 2\sum_{i=1}^{N} [\mu_i - n_i + n_{i}\ln{\frac{n_i}{\mu_i}}],
\end{equation}
where $n_i$ represents the number of events expected in bin $i$ for the ``truth'' model and $\mu_i$ represents the number for the test hypothesis in the same bin. If the test model has the same underlying parameters $m_X$ and $c\tau_X$ as the truth model then $\mu_i = n_i$ and $\chi^2\to 0$. Given that the predicted backgrounds of $e^+ e^-$ events in the DUNE ND-GAr are negligible~\cite{Berryman:2019dme}, we neglect these in calculating the $\chi^2$ test statistic.

Finally, we use this test statistic in two different ways to address the two questions posed above. When determining the parameter estimation capability within a given simplified framework (scalar case (\ref{eq:SimpleScalar}) or fermion case (\ref{eq:SimpleFermion}))
parameter space, we choose a truth $\{m_X,\ c\tau_X\}$ and then scan over the mass/lifetime within that model parameter space. We then estimate the allowed regions in the $m_X-c\tau_X$ plane by determining contours of $\Delta \chi^2$ with respect to the minimum corresponding to different confidence levels. For two-dimensional parameter measurement, we will present $\{1\sigma,\ 2\sigma,\ 3\sigma\}$ confidence-level regions, corresponding to $\Delta\chi^2 = \{2.3,\ 6.18,\ 11.83\}$. 

When attempting to determine how well we may discriminate between the scalar and fermion cases,
we compare a different quantity. For instance, if we assume that there is some 
scalar with mass $m_S$ and lifetime $c\tau_S$, we compare that distribution with test 
fermion
ones with some mass $m_N$ and lifetime $c\tau_N$. We scan over all possible $\{m_N,\ c\tau_N\}$, determining the minimum of the test statistic obtained in this process -- the larger this ``$\Delta \chi^2_{\rm model}$'' is, the easier it is to distinguish between the two model hypotheses.

\subsection{Mass and Lifetime Measurement Potential}\label{subsec:SimResults1}
\begin{figure}
    \centering
    \includegraphics[width=\linewidth]{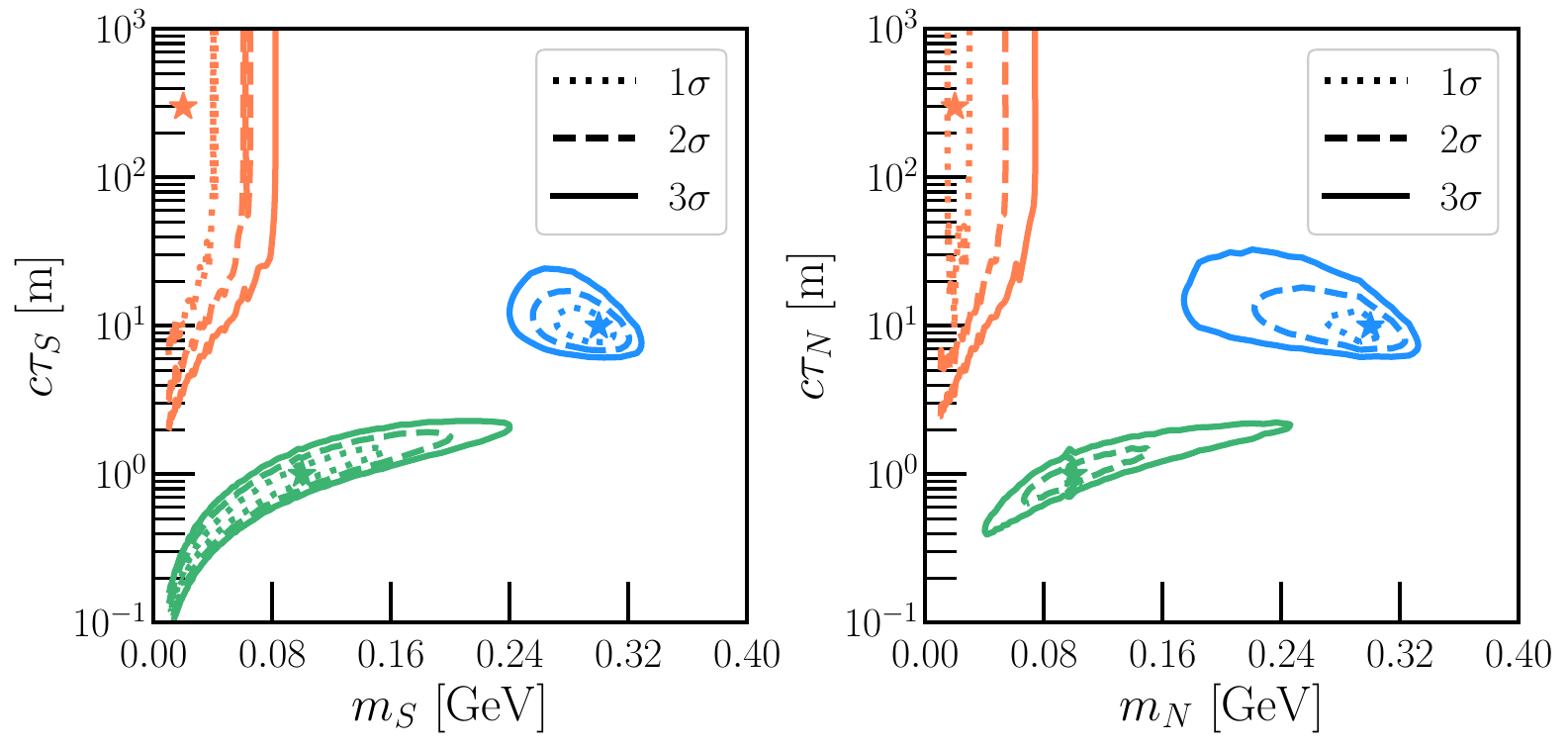}
    \caption{Measurement capability of the DUNE Gaseous Argon TPC to either a simplified framework scalar 
    $S$ (left panel) or 
    fermion
    $N$ (right), assuming the underlying model is known and that the data consists of 100 signal events. Different-color contours present the measurement capability for different combinations of the true new-physics particle mass and lifetime, and dotted/dashed/solid lines correspond to $1\sigma$/$2\sigma$/$3\sigma$ measurement expectations. See text for more detail.\label{fig:MeasurementPotential}}
\end{figure}

Given the details of Section~\ref{subsec:SimSetup}, we may now determine, if a given model is assumed, how well the underlying model parameters can be measured using data from the DUNE ND-GAr. We choose three different benchmark mass/lifetime points to serve as a representative sample: $\{m_X,\ c\tau_X\} = \{$100 MeV, 1 m$\}$, $\{$300 MeV, 10 m$\}$, and $\{$20 MeV, 300 m$\}$. We generate pseudodata for each of these points for the 
scalar case ($X = S$) and for the 
fermion case ($X = N$), and then fit the pseudodata to the model parameters $m_X$ and $c\tau_X$, assuming that the model is known. The results of this procedure are shown in Fig.~\ref{fig:MeasurementPotential} for the scalar 
(left) and 
fermion
(right) cases.

Some notable features from this analysis include
\begin{itemize}
    \item For sufficiently heavy, medium-lifetime particles (e.g. the blue contours in both panels of Fig.~\ref{fig:MeasurementPotential}), both the mass and lifetime of $X$ may be measured simultaneously. Despite imperfect detector resolution, the invariant mass of a fully visibly decaying scalar $S \to e^+ e^-$ is measured sufficiently well, and the energy distribution of the electron/positron pairs can be used to measure $c\tau_X$.
    \item If the particle is long-lived, $c\tau_X \gtrsim 100$ m, then only a lower limit on the lifetime may be placed. This behavior is evident for the orange contours, where the particle is very long-lived and light. Only a lower limit on the lifetime may be placed, and only an upper limit on the mass (due to imperfect measurement of $m_{e^+ e^-}$).
    \item For short-lived particles (e.g. the green contours), it is very challenging to measure the mass of the decaying particle. This is due to the fact that, for such short-lived particles to reach the detector and decay within, they must be highly boosted. Such boosted particles will decay into high-energy $e^+ e^-$ pairs which are more challenging to measure leading to less precision in determining $m_X$.
    \item Depending on the mass/lifetime combination, measurements are more precise for the 
    scalar 
    model in some situations and for the fermion
    in others.
\end{itemize}
The features present in both panels of Fig.~\ref{fig:MeasurementPotential} are general for different assumed-true combinations of $m_X$ and $c\tau_X$, for instance, if $m_S \approx$ 150 MeV and $c\tau_S \approx 30$ m, we would expect a modest determination of the scalar mass and only a lower limit on its lifetime. We have also explored the scenario in which the opening-angle information on $\theta_{e^+ e^-}$ is discarded in such an analysis (instead using only the invariant mass/energy of the $e^+ e^-$ pair). We find that the improvement with including $\theta_{e^+ e^-}$ is modest, helping especially for the case in which $m_X \gtrsim 100$ MeV, aiding in the measurement potential of $m_X$.

In the event of such a signal in a next-generation experiment, combinations of the event rate, the measured lifetime, and the measured mass can be used for validation of a specific model scenario. This is because, in specific models such as the Higgs-Portal Scalar and the Heavy Neutral Lepton, those three variables are not independent. Identification of a LLP that deviates from the model prediction would hint at a more complex dark sector than one of these specific models. 

It is worth emphasizing that in our simplified frameworks and analyses in this section, we have assumed the scalar or fermion dominantly decays to a final state containing an $e^+ e^-$ pair and that branching ratios to other possible channels are negligible. On the one hand, we note that it is possible to construct models where LLP decays to the $e^+e^-$ final state dominates (see discussion in Section~\ref{subsec:Lagrangian}), and more importantly, experiments such as DUNE can search for this final state for a broad range of LLP masses and naturally interpret the results of such searches within the simplified frameworks discussed here. On the other hand, additional decay channels can be present in particular UV model completions (e.g., in the Higgs Portal Scalar model, the scalar decay to muons will dominate over the one to electrons for $m_S > 2 m_\mu$). 
To account for such possibilities, it would be sensible to formulate and analyze simplified frameworks for a LLP decays to a variety of final states. We leave this to future work.

\subsection{Model Discrimination Potential}\label{subsec:SimResults2}
Now we turn to the question of model discrimination -- if a signal is present in the detector, how well can the underlying model 
(scalar vs. fermion cases, for example)
be determined. We quantify this by simulating data according to one of the models and a corresponding pair of mass/lifetime $m_{X}^{\rm true}$/$c\tau_{X}^{\rm true}$, then fitting that simulated data according to the \textit{other} model. We test all mass/lifetime pairs of the other model, finding the pair that yields the smallest test statistic $\chi^2$. Points in $\{ m_{X}^{\rm true}, c\tau_{X}^{\rm true} \}$ parameter space with a small test statistic in this procedure are those that are easier to confuse between the two models, and ones with a large test statistic will readily be identified as the correct model.

\begin{figure}
    \centering
    \includegraphics[width=\linewidth]{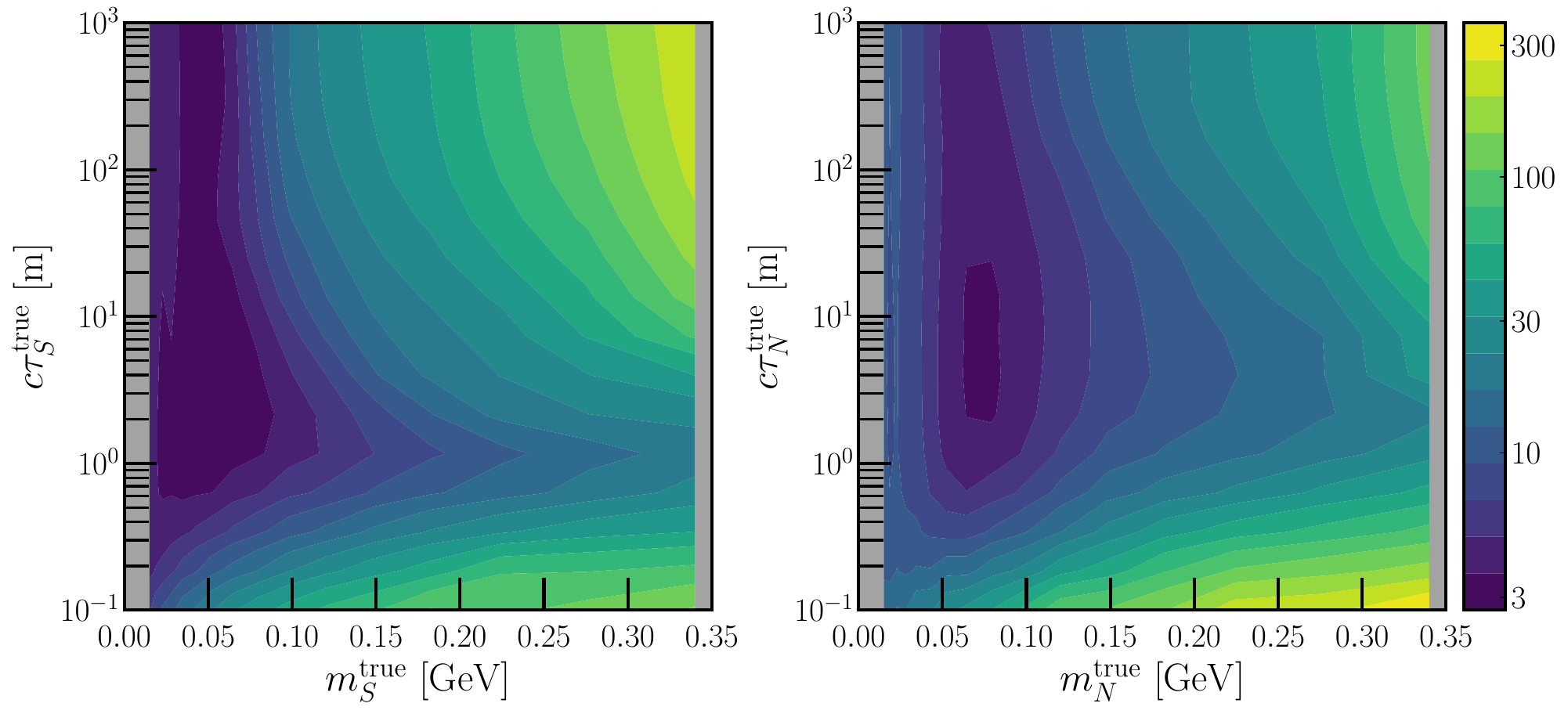}
    \caption{Minimum test-statistic $\chi^2$ when generated pseudodata according to a particular mass/lifetime in one new physics model and fitting the pseudodata to the \textit{wrong} new-physics model -- 
    simplified-framework scalar pseudodata (left) panel are fit according to the 
    simplified-framework fermion hypothesis (right) and vice versa. Darker regions of parameter space correspond to those where model discrimination is more challenging, and lighter ones correspond to points in truth parameter space where discrimination is more easily performed.}
    \label{fig:ModelDiscrimination}
\end{figure}

A contour plot of these minimum test statistics is shown in Fig.~\ref{fig:ModelDiscrimination}, where the true model is assumed to be the simplified framework scalar (fermion)
in the left (right) panel. Darker regions in these ``truth'' parameter spaces correspond to ``more confusing'' scenarios. For instance, in the left panel, when the true scalar mass is about 50 MeV and the lifetime about 3 meters, the minimum test statistic obtained when the simulated data is fit by the wrong model is about $\chi^2_{\rm min.} \approx 3$. Experimentally, with so many data points in such an analysis, it would be difficult to definitively conclude which underlying new-physics model is responsible for the data. In contrast, for a heavier mass around 300 MeV and a long lifetime around 100 meters or so, this minimum test statistic is significantly higher, $\chi^2_{\rm min.} \approx 200$ (note the logarithmic scale of the color bar). With such a case, the resulting data would much more obviously be fit by one model scenario over the other -- not only would new physics be discovered, but some identification of its source would be possible as well.

\begin{figure}
    \centering
    \includegraphics[width=0.8\linewidth]{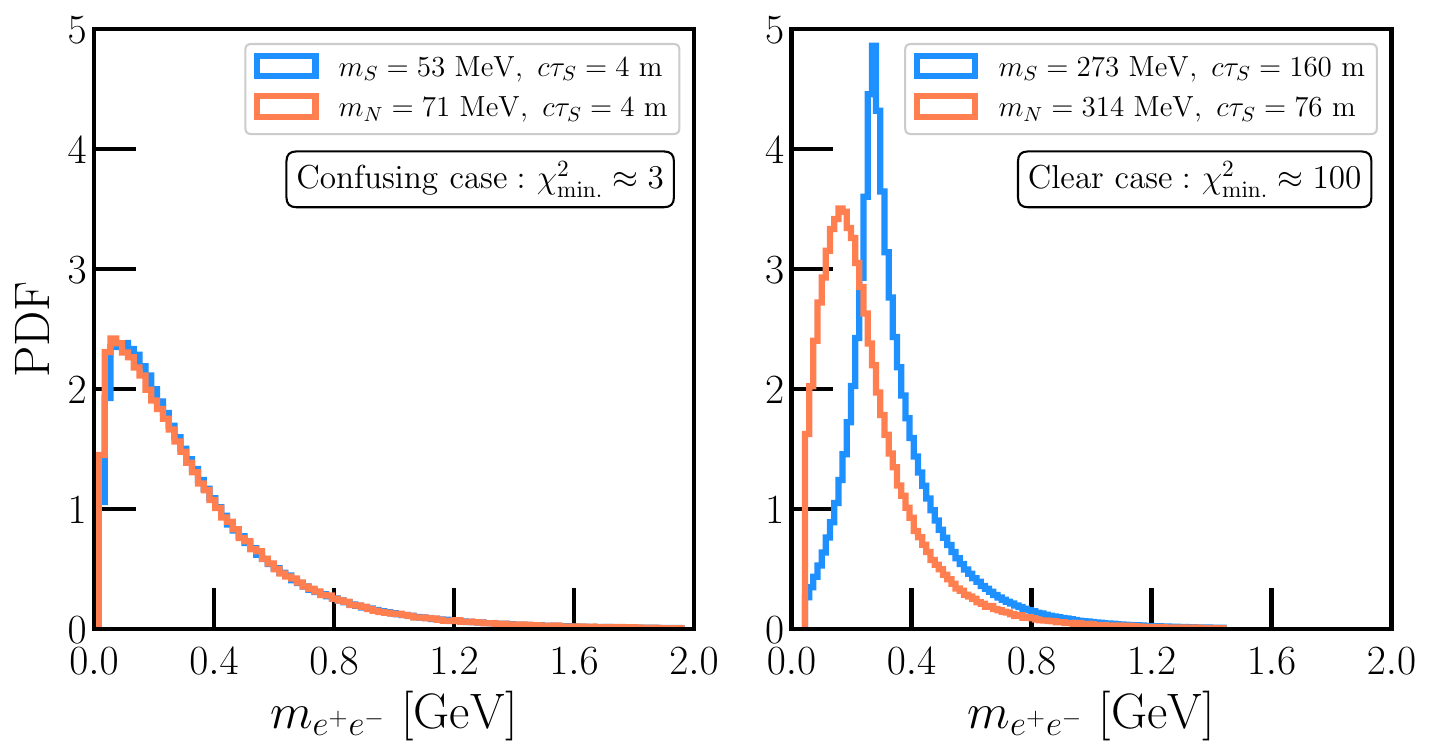}
    \caption{Kinematic distributions for scalar/fermion hypothesis testing in a situation in which the two are difficult (left panel) or easy (right panel) to disentangle. The combinations of $m_S$ and $c\tau_S$ are chosen according to Fig.~\ref{fig:ModelDiscrimination} and are compared against the combinations of $m_N$ and $c\tau_N$ for which the best fit according to the fermion scenario occurs. Labels show the value of the $\chi^2$ obtained at that best-fit point. See text for more detail.}
    \label{fig:ConfusingKinematics}
\end{figure}

To demonstrate this, Fig.~\ref{fig:ConfusingKinematics} presents a subset of the kinematics (the electron/positron invariant mass) expected in a ``confusing'' point of the (left) parameter space of Fig.~\ref{fig:ModelDiscrimination} and for a ``clear'' point. We show the kinematics according to the assumed-true parameters of the scalar $m_S$ and $c\tau_S$, as well as the kinematics for the fermion for which the best fit to that generated pseudodata is obtained, $m_N$ and $c\tau_N$. Fig.~\ref{fig:ConfusingKinematics}(left) shows this for a point where there will be great confusion, $m_S = 53$ MeV and $c\tau_S = 4$ m (the fermion fit prefers $m_N = 71$ MeV and $c\tau_N$ = 4 m). In contrast, the right panel shows a point where there will be clear model discrimination, $m_S = 273$ MeV and $c\tau_S = 160$ m (the fermion fit here prefers $m_N = 314$ MeV and $c\tau_N = 76$ m). The unseen parameters ($E_{e^+ e^-}$ and $\theta_{e^+ e^-}$) show qualitatively similar results -- near-overlapping curves for the ``confusing'' case and very different ones for the ``clear'' one.

As with the discussion in Section~\ref{subsec:SimResults1}, we have explored such a study here when we discard the information about the electron/positron opening angle $\theta_{e^+ e^-}$, relying instead on only the energy/invariant mass of the electron/positron pair. In this case, the two models become much more difficult to distinguish; for large swaths of true model parameter space, there exists a corresponding point in the wrong model's parameter space that reproduces the pseudodata nearly perfectly. This highlights the importance of the powerful direction measurement capabilities of a gaseous argon TPC, where ${\sim}5^\circ$ angular resolution provides the difference between demonstrating the existence of new physics and determining the nature of the new physics' signal.

The broad features of the two panels in Fig.~\ref{fig:ModelDiscrimination} are largely similar -- it is more challenging to determine the underlying model when the particle is light (between approximately 50 - 100 MeV) and with a lifetime around 10 meters. Heavier particles are easier to identify, especially when they are either long-lived (lifetime larger than about 10 meters) or very short-lived (lifetimes on the order of tens of centimeters and smaller). Comparing parameter spaces like this allows us to prepare for a potential discovery to benchmark expectations in the wake of a potential signal.

\section{Discussion \& Conclusions}
\label{sec:Conclusions}

Accelerator neutrino beam experiments are powerful tools in the search for new light weakly coupled physics. Most of the BSM searches at these experiments are driven by the ``top-down'' predictions in certain theoretically well motivated models. While this approach is sensible and should continue, it is evident that the specific final state signature in a given search may be realized in much broader range of UV models than the one that inspired the search. This observation motivates a more flexible theoretical framework in which to interpret the outcome of experimental searches. In this paper we have introduced simplified frameworks for BSM searches at neutrino beam experiments. These simplified frameworks may be specified by a small number of primary quantities, such as masses and lifetimes of particles, decay branching ratios, production and scattering cross sections, and production energy and position distributions, which directly determine the event rates and final state kinematics for the signature of interest. The results of experimental searches can be framed as constraints (in the event of a null result) or measurements (in the advent of an excess) of these primary quantities. Characterizing searches in this way will allow for straightforward reinterpretations in a variety of more complete theoretical constructions (simplified models, EFTs, and UV completions).

We have illustrated the approach by presenting a study of two concrete simplified frameworks for LLPs, one with a scalar LLP and the other with a fermion LLP, that are produced through kaon decays and decay to a final state containing an $e^+ e^-$ pair. These simplified frameworks are specified by just three parameters: the mass and lifetime of the LLP and a product of production and decay branching ratios which effectively fix the signal rate. We have illustrated how these simplified frameworks may be mapped to the well-studied Higgs-Portal Scalar or Heavy Neutral Lepton models, or alternatively to more elaborate theories with additional structure in the UV. We have derived the existing constraints on these simplified frameworks from a variety of experiments, including Super-Kamionkande, E949, NA62, MicroBooNE, as well as future projections for DUNE ND-GAr. We also showed how model-specific predictions are mapped to the simplified framework parameter space for the cases of the Higgs-Portal Scalar and Heavy Neutral Lepton models. As a byproduct of this investigation, we derived new leading constraints on the Higgs-Portal Scalar model from a search for Heavy Neutral Leptons at T2K. We also provided interpretations for a scenario in which a 100 event signal excess is observed at DUNE in the future. In this case, one can extract measurements of the simplified framework parameters such as LLP mass and lifetime, as well as distinguish one simplified framework from another. 

Looking ahead, it would be very interesting to formulate and analyze simplified frameworks for other signatures of interest at neutrino beam experiments. There are of course a variety of other proposed LLP scenarios involving different production mechanisms and decays. We suspect that one can straightforwardly formulate simplified frameworks for these scenarios, although there will be some interesting distinctions to those studied here, such as prompt production mechanisms (e.g., as with dark vector production through prompt neutral pion decays or proton bremsstrahlung) and different final states involving photons, muons, or hadrons. Another well-motivated scenario involves the production of dark matter and its subsequent scattering in the near detector~\cite{Batell:2009di,deNiverville:2011it,deNiverville:2012ij,Batell:2014yra,deNiverville:2015mwa,Coloma:2015pih,deNiverville:2016rqh,DeRomeri:2019kic,Dutta:2019nbn,Batell:2021blf}.
In developing a simplified framework description for this case, one potential challenge is to devise a minimal parameterization of the cross section which adequately captures the kinematics of the scattered final state particle. Similar considerations may apply to simplified frameworks for inelastic dark matter~\cite{Izaguirre:2017bqb,Jordan:2018gcd,Berlin:2018jbm,Tsai:2019buq,Batell:2021ooj}, which feature LLP decays and up- and down-scattering signals, as well as those which characterize neutrino-induced BSM signals, such as dark neutrinos which up-scatter in the near detector to a dark fermion~\cite{Gninenko:2009ks,Bertuzzo:2018itn,Ballett:2018ynz,Arguelles:2018mtc,Kamp:2022bpt,Abdullahi:2022cdw,Abdallah:2020biq,Abdallah:2020vgg}.

Generally, such simplified frameworks offer a powerful way to analyze current and upcoming data in a very broad way, allowing us to search for new physics in a broad, yet efficient, manner. They allow for searches that can set powerful constraints on parameter space, or better yet, to characterize a detected signal in the presence of new physics. Long-lived particle searches at these neutrino facilities are impressive sites for new-physics searches, and providing tools such as these simplified frameworks will maximize our chances of such discoveries in the decades to come.

\begin{acknowledgments}
The work of B.B. and W.H. is supported by the U.S. Department of Energy under
grant No. DE–SC0007914. This work was performed in part at Aspen Center for Physics, which is supported by National Science Foundation grant PHY-2210452.
\end{acknowledgments}

\bibliographystyle{JHEP}
\bibliography{refs}

\end{document}